\documentclass[useAMS,usenatbib]{mn2e}
\usepackage{graphicx}
\usepackage{graphics}
\usepackage{epstopdf}
\usepackage{float}
\usepackage{bm}
\usepackage{epsf}
\usepackage{url}
\usepackage{natbib}
\usepackage{amsmath}
\usepackage[longend]{algorithm2e}
\usepackage{color}

\newcommand{\rev}[1]{\textcolor{black}{#1}}

\newcommand{\mpc}{{h$^{-1}$Mpc \;}}

\title[Hierarchical Reconstruction of the Cosmic Web]{Hierarchical Reconstruction of the Cosmic Web, The H-Spine method.}
\author[Aragon-Calvo M.A. et al.]{M.A. Aragon-Calvo$^{1}$\thanks{E-mail:maragon@astro.unam.mx}\\
$^{1}$Instituto de Astronomía, UNAM, Apdo. Postal 106, Ensenada 22800, B.C., México.\\}
\begin{document}

\date{}

\pagerange{\pageref{firstpage}--\pageref{lastpage}} \pubyear{2002}
\maketitle
\label{firstpage}

\begin{abstract}

The cosmic web consists of a nested hierarchy of structures: voids, walls, filaments, and clusters. These structures interconnect and can encompass one another, collectively shaping an intricate network.
Here we introduce the Hierarchical Spine (H-Spine) method, a framework designed to hierarchically identify and characterize voids, walls, and filaments. 
Inspired by the geometrical and dynamical constraints imposed by anisotropic gravitational collapse, the H-Spine method captures the geometry and interconnectivity between cosmic structures as well as their nesting relations, offering a more complete description of the cosmic web compared to single-scale or multi-scale approaches. 

To illustrate the method's utility, we present the distribution of densities and sizes of voids, walls and filaments identified in a 3-level hierarchical space. This analysis demonstrates how each level within the hierarchy unveils distinctive densities and scales inherent to cosmic web elements.

\end{abstract}

\section{Introduction}

The distribution of galaxies in the Universe, along with their underlying density field, form an interconnected web of structures spanning a wide range of scales and densities. This complex pattern, revealed in early galaxy maps \citep{Joeveer78,Tarenghi78,Einasto80,Tully87,Geller89} and later confirmed in large redshift surveys \citep{Lapparent86,York00,Jarrett03,Courtois13} is the result of the gravitational collapse of primordial fluctuations.  
The growth of structures in the universe is driven by the hierarchical gravitational collapse of matter  \citep{PressSchechter74,White78}, in which small primordial density fluctuations, originated during the inflationary epoch, grew and merged to form more massive structures. Within this framework, density fluctuations experienced a sequential collapse: initially on the smallest levels and subsequently aggregating into galaxies, groups, and clusters of galaxies \citep{White91,Lacey93,Navarro95,Cole00}.
The above picture, however,  is incomplete: the gravitational collapse of matter is inherently anisotropic. A cloud of matter initially collapses to form a wall, then a filament, and finally a node \citep{Zeldovich70}. The Large Scale Structure (LSS) is the result of the gravitational collapse of matter occurring at different scales, producing in a network of interconnected clusters, groups, and galaxies linked by filaments that form a mesh of walls, which in turn enclose cosmic voids \citep{Bond96,Weygaert02}. The multiscale nature of the primordial power spectrum gives rise to a hierarchy  in which voids contain sub-voids which themselves define an inner network of walls, filaments and nodes; walls contain sub-walls that define a network of filaments and nodes and filaments contain sub-filaments defining nodes.

The study of the geometry, dynamics and nesting relations between the elements of the cosmic web is fundamental for a complete understanding of the formation and evolution of both the LSS and the galaxies within. The cosmic web, being the stage where galaxy evolution takes place, has a directly effect on galaxy formation processes, as shown in the morphology-density relation \citep{Dressler80}, the spin alignment of galaxies in walls and filaments \citep{Patiri06,Aragon07Spin, Hahn07}, the low velocity dispersion of galaxies inside walls \citep{Aragon11} and the stripping of gas as galaxies interact with their host wall \citep{Benitez13} to name a few. The evolution of the cosmic web being affected by the background cosmology can also be used as a testing probe, as in the case of voids, which are particularly sensitive to dark energy \citep{Pisani15,Pollina16,Bos12}. Cosmic structures can also potentially point to changes in the cosmological model, for instance, recently \citet{Peebles23} found a large wall extending from the Local Supercluster (see also \citep{Peebles22}) which may be in tension with current models of structure formation. In order to explain and interpret these and other observations we need detailed and quantitative descriptors of the LSS.

\subsection{Cosmic Web finders}

The identification and characterization of the elements of the cosmic web is a challenging task. The wide range of scales involved, high dynamic range in densities and the non-trivial relation between elements of the distribution of matter prevents the use of simple prescriptions such as density thresholds. Many approaches have been proposed to identify and characterize the Large Scale Structure.  Although a division of cosmic web classification methods by type is somewhat artificial it is nevertheless useful for our discussion. Cosmic web finders can be divided according to the nature of their algorithms into 
local gradient, topology, point-processes and, recently, machine learning. The following is a non-exhaustive discussion of different methods presented in the literature. For a more complete review of the general state of the field we refer the reader to the cosmic web finder comparison project \citep{Libeskind18}.

\subsubsection{Local gradient methods}

Local gradient methods associate local variations in the density field to a set of basic morphologies quantified by means of the Hessian matrix \citep{Aragon07b,Hahn07}. The morphological analysis is performed either at some fixed smoothing length or in scale-space. Local gradient methods have been extended to include other scalar and vector fields \citep{Forero09,Hoffman12,Cautun13}. One of the major shortcoming of local gradient methods is their reliance on a global thresholding operation applied to the filtered field. This threshold is determined empirically to produce the required segmentation\footnote{We refer to segmentation as the labeling of different regions in the density field according to some creiteria.} and often results in discontinuous structures in noisy/low contrast regions. Gradient-based methods produce a voxel-wise segmentation of the LSS and are unable to identify individual structures or produce catalogs of structures, although there have been efforts to model filaments and walls after post-processing the segmentation masks \citep{Aragon10c,Cautun14,Pfeifer22}.

\subsubsection{Filament finders}

Given the prominence of filaments in the distribution of galaxies a significant effort has been made to develop specific methods to identify them in both galaxy surveys and simulations. Two popular methods based on point processes are the Candy and Bisous filament finders \citep{Stoica05,Tempel16} which identify filaments by associating galaxies inside cylinders. Other methods include the graph theory-based T-rex filament finder \citep{Bonnaire20} and the 2D sky-based filament finder presented in \citet{Duque22}. However many of these methods lack a clear physical motivation and/or are designed for niche applications.

\subsubsection{Machine learning methods}

In recent years the field of cosmic web analysis has benefited from advances in machine learning techniques. These new methods cover a wide range of techniques with the common theme of training a system based on labeled data \citep{Aragon19, Inoue22, Carron22}. These methods are highly effective and efficient, often outperforming traditional methods at a fraction of the computational cost. However, they lack physical motivation and interpretability, being in effect a black box. In addition to this, machine learning methods can be brittle and underperform when the input dataset is different from the training set. The difference between the input and training datasets can be small and in some extreme cases even indistinguishable \citep{Chakraborty18}. This limits their applicability and to date these methods are regarded more as experimental applications and demonstration of the capabilities of the technology than real workhorse tools.

\subsubsection{Topology-based methods}

Topological cosmic web finders, like the one presented here, are based on the critical points of the density field sampled on a regular grid such as the Spine method \citep{Aragon10b} or irregular point distribution such as the ZoBov, Disperse and Origami methods \citep{Neyrinck08,Sousbie11,Falck12}. In topology-based methods the critical points in the density field are associated to the morphological elements of the cosmic web, taking advantage of the correspondence between the topology of the density field and its geometry. Topological methods are a separate class of structure finders in their ability to not only classify the LSS into its basic morphologies but also being able, by construction, to identify individual structures. These methods are also insensitive to local variations that can fragment structures as in the case of local gradient methods. However, they are prone to \textit{oversegmentation}, which is the artificial segmentation of structures due to small noisy structures that can change the topology of a region. In order to alleviate this problem a threshold is usually applied \citep{Neyrinck08,Sousbie11} but this adds one free parameter with no direct physical interpretation. The Spine method relies on a smoothing of the density field to reduce oversegmentation in which the smoothing length is a free parameter. This can affect the shape of the identified structures if a large Gaussian smoothing kernel is used. 

\subsubsection{Hierarchical descriptions of the cosmic web}

Studies of the cosmic web have have, either directly or indirectly, uncovered some aspects of its hierarchical nature. The use of thresholds to select structures at different levels of saliency has been used by several authors \citep{Schmalzing99,Neyrinck08Zobov,Sousbie11,Weygaert13Alpha} and can be seen as a first approximation to extract hierarchical relations, although indirectly and not explicitly stated. \citet{Hahn07} noted that when increasing the smoothing length used to compute the morphology of the matter distribution the dimensionality of the structures also changed, indirectly uncovering the hierarchical relation between them. \citet{Aragon14} found that the strength of the spin alignment of a galaxy depends on the location of its parent structure within a hierarchy of structures.  \citet{Alpaslan14} found thin filaments crossing the interior of voids, indicating that voids, far from being empty and simple regions, contain a rich hierarchy of inner structures. This hierarchy was explicitly studied by \citet{Aragon10b,Aragon13a} although  restricted to voids. To date there has not been a characterization of the cosmic web that include both geometry and explicit hierarchical relations. The work presented here is a step forward a full characterization of the LSS into voids, walls and filaments in a hierarchical fashion.

\subsection{Anisotropic gravitational collapse}\label{sec:grav_collapse}

The elements of the cosmic web we observe in the distribution of matter are the result of fluctuations imprinted in the primordial density field. These initially shallow fluctuations grew as the result of the anisotropic gravitational collapse \citep{Zeldovich70} as:

\begin{equation}
\frac{\rho(x,t)}{\rho} \propto  \frac{1}{  (1- a(t)\lambda_1) \; ( 1-a(t)\lambda_2) \;( 1-a(t)\lambda_3) }
\label{eq:zeldovich}
\end{equation}

\noindent where $\lambda_3 > \lambda_2 > \lambda_1 $ are the eigenvalues of the deformation tensor. Gravitational collapse occurs when at least one of the eigenvalues is positive. The Zel'dovich formalism provides a description of the the gravitational evolution of a cloud of matter as a sequence of partial collapses: first into a wall, then a filament and finally a halo/cluster (see \citet{Shen06} and \citet{Hidding14} for useful insights). The collapse of a cloud of matter results in the reduction of dimensionality in the original cloud from a volume/void, to a sheet/wall, a line/filament and finally a point/halo (see table \ref{tab:zeldovich_collapse}). The Zel'dovich collapse encoded in eq. \ref{eq:zeldovich} is an idealized approximation. In reality gravitational collapse is a far more complex process modulated by the expansion of the Universe and the location of the cloud with respect to its surrounding structures.

\begin{table}
\begin{tabular}{ l c c c c c c }
\hline
Structure & \# dims & \# collapses & $\lambda_1$ & $\lambda_2$ & $\lambda_3$ & expansion \\ \hline
Void      & 3    & 0            & $\ominus$   & $\ominus$   & $\ominus$   & D$^3$ valley \\
Wall      & 2    & 1            & +           & $\ominus$   & $\ominus$   & D$^2$ valley \\
Filament  & 1    & 2            & +           & +           & $\ominus$   & D$^1$ valley \\
Node      & 0    & 3            & +           & +           & +           & \\ \hline
\end{tabular} 
\caption{Number of dimensions (\# dims) and number of gravitational collapses (\# collapses) leading to the four basic morphological types determined by the eigenvalues of the deformation tensor. The last column shows the number of dimensions in which a structure with negative eigenvalues expands.}\label{tab:zeldovich_collapse}
\end{table}

\begin{figure}
	\centering
	\includegraphics[width=0.4\textwidth]{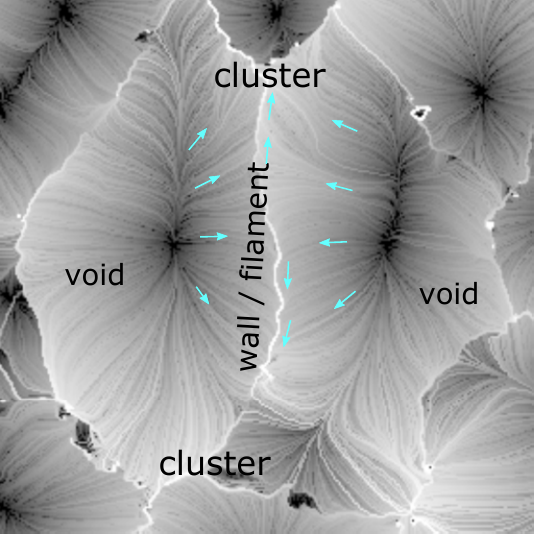}
	\caption{Velocity field around a wall/filament located between two voids. The two adjacent voids expand, driving matter into their common  wall/filament at the center of the figure. The central wall/filament drives matter into the nearby clusters located at the center top and bottom of the figure. The direction and sense of the velocity field is shown with blue arrows and using an advection technique (see \citet{Aragon13Hierarchy}).}
	\label{fig:LIC}
\end{figure}

\begin{figure*}
	\centering
	\includegraphics[width=0.99\textwidth]{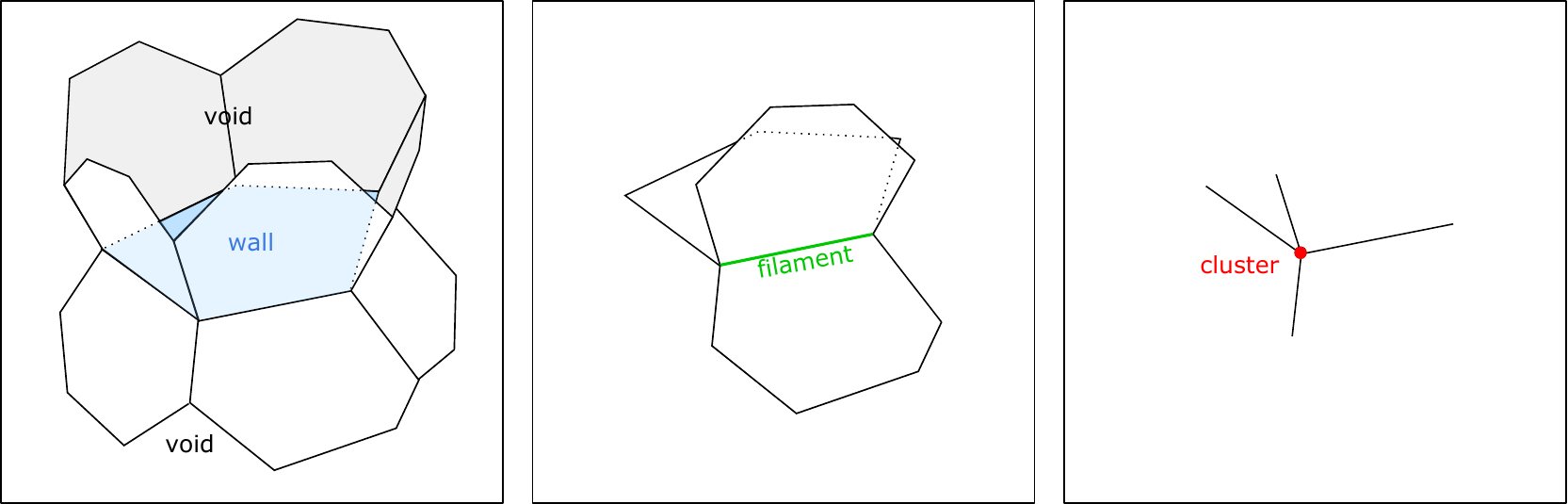}
	\caption{Connectivity of the basic elements of the cosmic web. Walls are located at the intersection of voids. Filaments are located at the intersection of walls and clusters are located at the intersection of filaments. Note that all elements of the cosmic web are interconnected.}
	\label{fig:cosmic-web-connectivity}
\end{figure*}

\subsection{Connectivity of the cosmic web}\label{sec:web-connectivity}

The connectivity of the elements of the cosmic web, although not explicitly stated, can be inferred from the Zel'dovich collapse. Voids, having three negative eigenvalues, expand in three dimensions, driven by the collapse of its surrounding walls, filaments and nodes. Walls, with two negative eigenvalues, expand in two dimensions driven the the collapse of its surrounding filaments and nodes. Filaments, having one negative eigenvalue expand in one dimension driven by the collapse of its adjacent nodes (see table  \ref{tab:zeldovich_collapse}). Adjacent expanding voids define the locations of collapsing walls, adjacent expanding walls (along their plane) define the location of collapsing filaments and adjacent expanding filaments (along their main axis)  define the location of collapsing clusters. 
In this picture, walls are located at the intersection of pairs of voids, filaments are located at the intersection of walls and clusters are located at the intersection of filaments as illustrated in Fig. \ref{fig:cosmic-web-connectivity} (see also \citet{Weygaert94}).

\begin{table}
\begin{tabular}{ l c c c c }
\hline
          &       & D$^3$ & D$^2$ &  D$^1$ \\
\hline
Voids     & $\to$ & sub-voids & sub-walls & sub-filaments \\
Walls     & $\to$ &           & sub-walls & sub-filaments \\
Filaments & $\to$ &           &           & sub-filaments \\
\end{tabular}
\caption{Inner hierarchy of structures divided by morphological type. Structures of dimension D$^n$ can contain only sub-structures of dimension D$^{\le n}$.}
\label{tab:spine_hierarchy}
\end{table}

\begin{figure}
	\centering
	\includegraphics[width=0.4\textwidth]{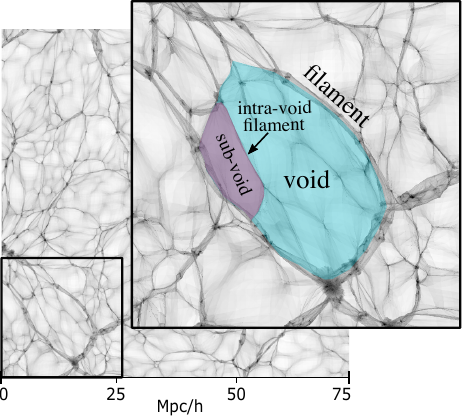}
	\caption{\rev {Hierarchical relations in the cosmic web, here highlighting a void and its internal structure. The highlighted blue region corresponds to a void delineated by large an dense filaments. Inside this void there are several sub-voids (one highlighted in purple) delineated by intra-void filaments (note that this is a thin 2D slice. In the full 3D case voids are delineated by walls)}.}
	\label{fig:cosmic-web-hierarchy}
\end{figure}

\subsection{The hierarchical cosmic web}

In the standard model of structure formation haloes aggregate into larger structures to form galaxies, groups and clusters \citep{Bardeen86,Bond91,Lacey93,Sheth01}. The full formation history of a collapsed object is already present in the initial conditions as a hierarchy of fluctuations at different scales whose relations are encoded in the power spectrum. In this picture the peaks in the primordial density field contain all the information describing the hierarchy of smaller peaks. At the opposite extreme in density the underdense voids contain a hierarchy of smaller voids which themselves can contain an inner hierarchy of even smaller voids, walls, filaments and nodes \citep{Gott03,Sheth04,Aragon10b,Paranjape12,Neyrinck13}. Similarly, walls contain an inner hierarchy of smaller walls, filaments and nodes. Filaments contain an inner hierarchy of smaller filaments and nodes (see \citet{Cadiou20} for a scale-space analytic approach).

\subsubsection{Dimensionality restrictions}

The geometry and connectivity of the cosmic web impose geometrical restrictions in the hierarchy:  an element of the cosmic web of dimension D$^n$ can only contain an inner hierarchy of elements of dimension D$^{\le n}$ (see table \ref{tab:spine_hierarchy}). \rev{Figure \ref{fig:cosmic-web-hierarchy} illustrates this scenario by showing a void delineated by a shell of walls and filaments and a sub-void defined by a smaller shell of walls and filaments, these structures being inside the larger void. }

\section{The new Spine formalism}\label{sec:spine}

Here we briefly describe the original Spine method and introduce its new formalism. For a more detailed description of the Spine method we refer the reader to \citet{Aragon10}. The Spine formalism extends the Watershed Void Finder (WVF) introduced by \citet{Platen07} based on the watershed transform method \citep{Gauch99}. The WVF identifies voids as regions in the density field that share the same local minima. It produces a segmentation of space into contiguous regions each corresponding to an individual void. The Spine formalism identifies walls, filaments + nodes (called the spine) as follows:

\begin{itemize}
\item Walls $\to$ intersection of two \textit{voids}.
\item Spine $\to$ intersection of three or more \textit{voids}.
\end{itemize}

\subsection{New Spine formalism}

The Spine prescription described above, while simple and effective, is not consistent with the dimensional relations and intrinsic connectivity of the elements of the cosmic web (see Tables \ref{tab:zeldovich_collapse} and \ref{tab:spine_hierarchy}). The order in the gravitational collapse provides a natural basis to identify voids, walls and filaments as follows:

\begin{itemize}
\item Walls $\to$ intersection of two \textit{voids}.
\item Filaments $\to$ intersection of two or more \textit{walls}.
\item Nodes $\to$ intersection of two or more \textit{filaments}.
\end{itemize}

\noindent This new Spine formalism is illustrated in Fig. \ref{fig:cosmic-web-connectivity}. The above prescription must be performed sequentially: first identifying voids then walls and finally filaments. As in the original Spine method we do not attempt to identify nodes since the connectivity of voxels in a regular grid (26 neighbours around a given voxel) can not adequately  describe the topology around a node except in the simplest cases. In practice this does not represent a problem since one can easily identify groups using algorithms such as Friends of Friends.

\begin{figure*}
	\centering
	\includegraphics[width=0.99\textwidth]{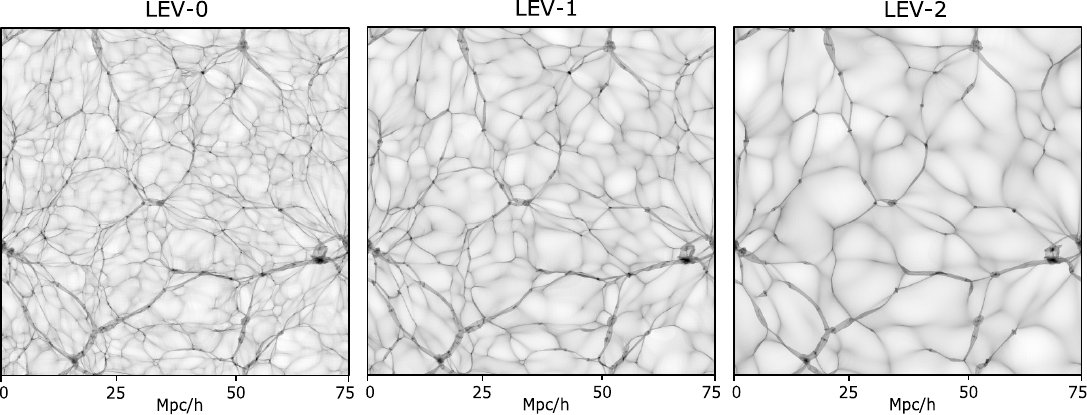}  
	\caption{Density field across the TNG100 simulation box showing the three levels of the hierarchical space used in this work. From left to right we show the $z=0$ density field evolved from initial conditions sharp-$k$ smoothed at 1, 2 and 4 \mpc respectively. For clarity we show the inverse of the density field with arbitrary scaling and clipped at $\delta + 1 < 200$.}
	\label{fig:density_hierarchical_space}
\end{figure*}

\section{Hierarchical Spaces}\label{sec:hierarchical-spaces}

One of the key aspects of the method presented here is the identification of structures in hierarchical space. We define a hierarchical space as a generalization of a scale space in which the different levels of the space encode both \textit{scale} and \textit{hierarchical relations} instead of just scale as in standard scale spaces \citep{Aragon10b,Aragon13a}. 

\subsection{Constructing hierarchical spaces: initial conditions smoothing}

The hierarchical nature of the cosmic web, imprinted in the initial conditions, allows us to construct a hierarchical space. The hierarchy in the cosmic web is exposed by removing increasingly larger scales in the initial conditions.\footnote{See the void formalism of \citet{Sheth04} } In practice this is done with a sharp$-k$ filter. This approach was first introduced by \citet{Einasto11}  in order to study the density fluctuations responsible for the emergence of cosmic voids. We construct a hierarchical space by sharp$-k$ filtering an initial conditions file at several scales. An ideal scale$/$hierarchical space is continuous, in practice a set of discrete scales are selected based on some criteria. Following \citep{Sato98} we select discrete scales as: $\sigma_k = S^n$, where $\sigma_k$ is the smoothing length to be applied to the initial conditions (in $k$-space), $S$ is a scaling factor and $n$ is the level of interest. The scaling factor determines the spacing between scales. \citet{Sato98} found that values in the range $1.5 < S < 2$ produce results within a few percent from the continuous case\footnote{\rev{Although it is important to note that their results are based on a regular scale space}}. Note that while a sharp$-k$ filter would be inappropriate for an evolved non-linear density field, in our case it introduces no artifacts due to the independence between Fourier modes in the initial conditions.

On the other hand, a scale-space constructed by Gaussian smoothing an evolved density field acts on a highly non-linear density field, further mixing Fourier modes that have been already mixed by gravitational collapse. Gaussian-based scale-spaces also isotropize the density field, smearing out the anisotropic features in the cosmic web we seek to identify. In sharp contrast an initial-conditions smoothed simulation lacks small scale primordial fluctuations that would produce sharp small-scale features such as dense, compact low-mass haloes, while keeping only the large-scale structure above the smoothing scale. The evolved density field produced by initial conditions smoothing results in a smooth version of the cosmic web that retains the anisotropic features of the original.

Figure \ref{fig:density_hierarchical_space} shows a hierarchical space consisting of three levels corresponding to a sharp-$k$ smoothing equivalent of 1, 2 and 4 \mpc, denoted LEV-0, LEV-1 and LEV-2 respectively. For simplicity we did not include larger scales in the hierarchy. 
Note that while the increasingly larger linear-regime smoothing erases small-scale structures it keeps the anisotropy and sharpness of the large-scale structures.
The structures in the linear-regime smoothed versions are also closer to their idealized shapes as in the case of the large voids in LEV-2 which appear as smooth underdense regions with clear borders and simple internal density gradients. In contrast their LEV-0 counterparts contain a complex network of sub-structures, their borders are not so clearly defined and their density profile is not evident. Similarly filaments in LEV-2 appear as smooth bridges joining clusters. Filaments in LEV-0 on the other hand, contain complex substructure including smaller filaments and several low-mass haloes.

\begin{figure}
  \centering
  \includegraphics[width=0.48\textwidth,angle=0.0]{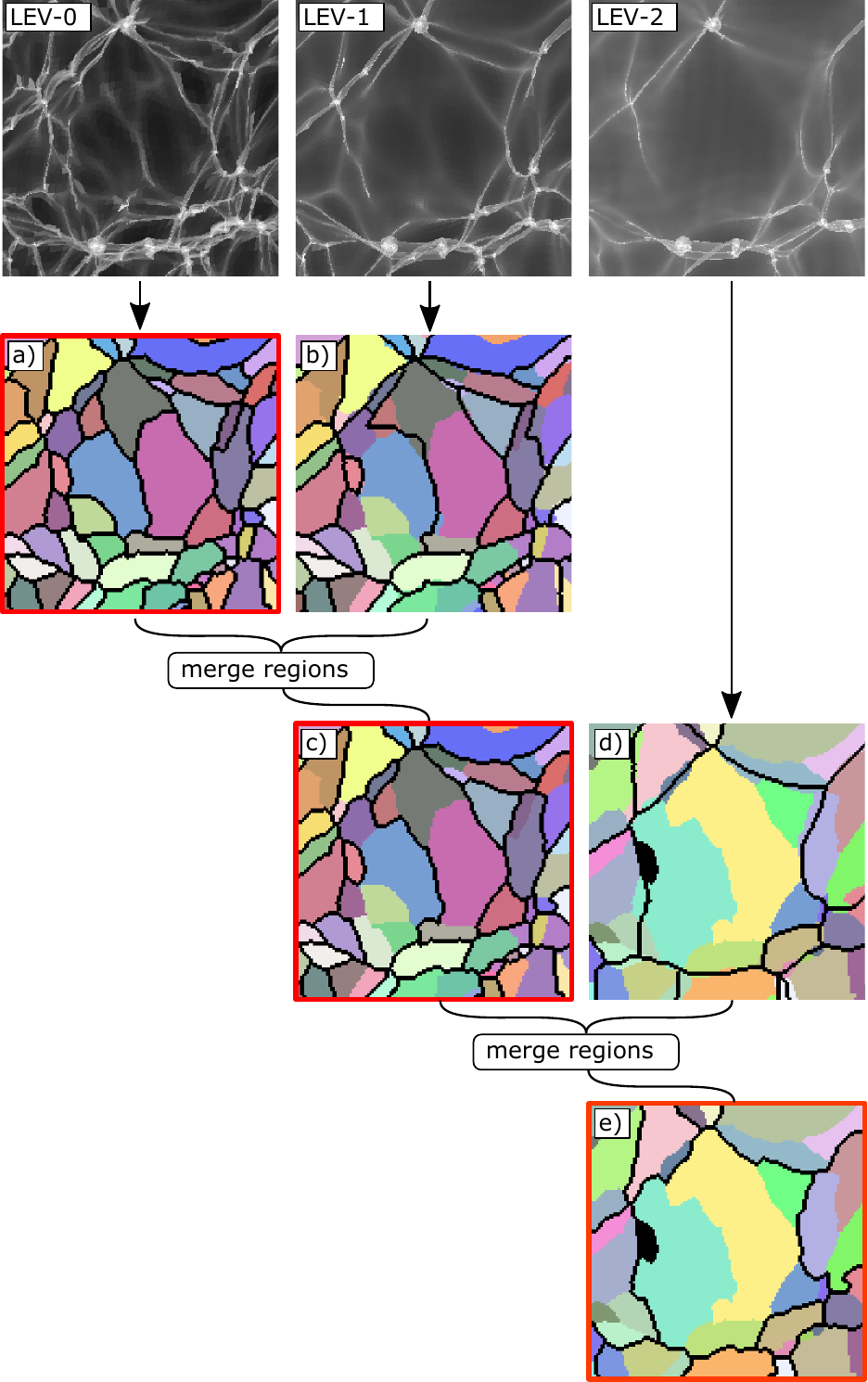}
  \caption{Recursive void merging across a 3-level hierarchical space. For clarity we show the results of a 2D simulation. The top panels show the density field smoothed at three different scales. Panel a) shows the watershed transform applied to LEV-0. Panel b) shows the watershed regions in LEV-1 superimposed on the incomplete watershed regions in LEV-0. Note the mismatch between regions in LEV-1 and LEV-0. In panel c) we merge the children regions in LEV-0 into the regions in LEV-1 as described in Sec. \ref{sec:h-spine}. This procedure is repeated between regions in LEV-2 and LEV-1 (panels c) y d) respectively) producing the reconstructed regions in LEV-3 shown in panel e). The reconstructed regions a), c) and e) are highlighted by a red box. This merging process can be iteratively repeated for any number of hierarchical levels. Note that the reconstructed regions follow the structure in the density field of LEV-0. For clarity we show the watershed boundaries, although the merging process is done with the incomplete watershed transform.}
  \label{fig:void_merge}
\end{figure}

\noindent We note that sharp-$k$ initial conditions smoothing is not the only or best way of producing a hierarchical or scale space. There are other approaches such as  isotropic$/$anisotropic smoothing \citep{Aragon07b,Hahn07,Cautun13}, a simple thresholding which can be implemented directly on the density field \citep{Platen07} or as a post-processing step \citep{Sousbie11, Neyrinck11ZOBOV}. Each of these approaches may have advantages over sharp-$k$ initial conditions smoothing in some particular applications.

\section{Hierarchical-Spine (H-Spine)}\label{sec:h-spine}

\begin{figure}
  \centering
  \includegraphics[width=0.3\textwidth,angle=0.0]{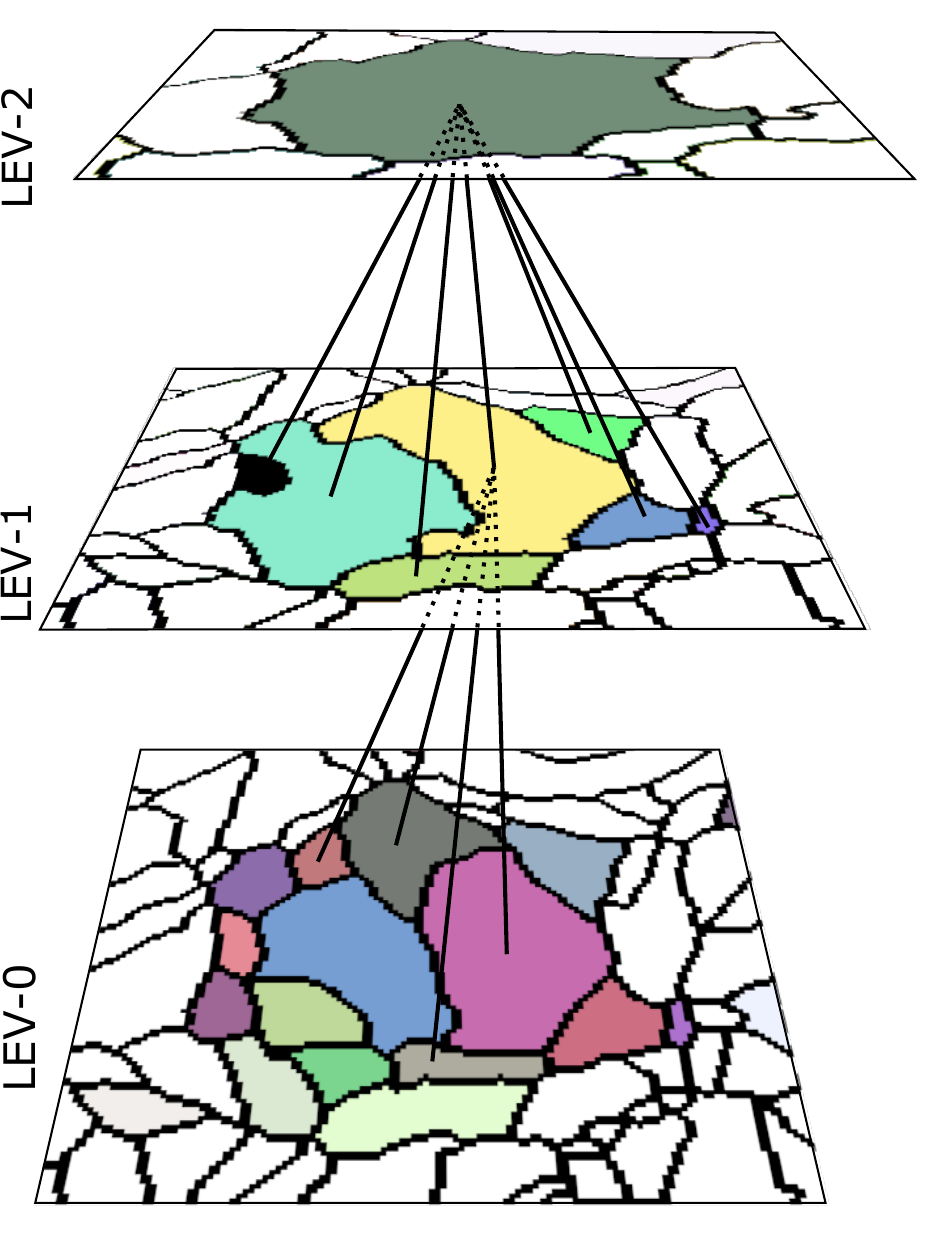}
  \caption{Hierarchical relations in a 3-level hierarchical space. We show a large void at LEV-2, its subvoids at LEV-1 and sub-sub voids at LEV-0. The lines indicate dependency between structures at adjacent levels in the hierarchy. For clarity we only show lines between one void at LEV-1 and its children voids at LEVEL-0.}
  \label{fig:void_merge-tree}
\end{figure}

In this section we describe the hierarchical implementation of the Hierarchical Spine method (H-Spine): 

\begin{itemize}

\item We begin by constructing a set of initial conditions forming a hierarchical space, as described in Sec. \ref{sec:spine}. Each of the initial conditions, corresponding to a hierarchical level, is evolved with an N-body code as described in Appendix \ref{sec:hspace_n_body}. 

\item From the evolved particle distribution we compute density fields interpolated on a regular grid.

\item From the density field in each level of the hierarchy we identify void regions by computing the incomplete watershed transform, i.e. watershed regions with no watershed boundaries, see \citet{Aragon13Hierarchy} for details. Each region of the incomplete watershed transform corresponds to a void. 

\item At this point we have voids identified at each level in the hierarchy. In the next step we merge voids between adjacent levels in the hierarchy. We follow the prescription presented in \citet{Aragon13Hierarchy}. We establish relations between voids at adjacent levels as follows: a parent void at level $i$ is defined by its children voids at level $i-1$. We associate voids in adjacent levels of the hierarchical space by identifying overlapping voxels. A given children void usually shares volume with several parent voids higher in the hierarchical space and we associate a given child with the parent with which it shares most of its voxels. \footnote{Note that this procedure is identical to the way halo merger trees are constructed} A void at level $i$ is then reconstructed as the union of its children voids at level $i-1$ as:

\begin{equation}\label{eq:merging}
\mathcal{V}_p = \bigcup v_c
\end{equation}

\noindent where $\mathcal{V}_p$  is the set of voxels defining a parent void and $v_c$ are the voxels defining each child void contained by $\mathcal{V}_p$, see Fig. \ref{fig:void_merge}.

\item After the merging procedure we have a hierarchical space in which voids at some level in the hierarchy can be recursively reconstructed by voids at lower levels. In Fig. \ref{fig:void_merge-tree} we show the hierarchical relations between voids at different levels in the hierarchy. Voids in LEV-2 are defined by voids in LEV-1 which are reconstructed in terms of voids in LEV-0. 

\item From the hierarchically-merged watershed regions we identify voxels at the boundaries between adjacent voids (i.e. the watershed transform) at each level. We then iterate over all the voxels in the watershed transform and identify voxels in walls following the criteria described in Table \ref{tab:spine_hierarchy}. We generate wall catalogs by grouping voxels sharing the same pair of adjacent voids .

\item We iterate again over the remaining voxels of the watershed transform and identify voxels in filaments following the criteria described in Table \ref{tab:spine_hierarchy}. We generate filament catalogs by grouping voxels sharing the same pair (or n-plet) of adjacent walls. 

\end{itemize}

The H-Spine pipeline described above can be summarized as follows: 

\begin{itemize}
\item[1)] Generate set of smoothed ICs (hierarchical space).
\item[2)] Evolve set of ICs to desired redshift.
\item[3)] Compute density fields.
\item[4)] Compute incomplete watershed regions for all hierarchical levels.
\item[5)] Hierarchically merge void regions between levels.
\item[6)] Complete watershed transform for all levels.
\item[7)] Compute Spine (voids, walls and filaments).
\item[8)] Compute Void catalog.
\item[9)] Compute Wall catalog.
\item[10)] Compute Filament catalog. 
\end{itemize}

The H-Spine method consists of 6 programs written in c and one python script that orchestrates the pipeline. The watershed and hierarchical merging codes are improved versions of the codes presented in \citet{Aragon10} and \citet{Aragon13a}. Additionally, the H-Spine pipeline includes new codes for computing wall and filament catalogs using the new Spine formalism described here. The full pipeline can process a 3-level hierarchy of $512^3$ voxel datacubes in a few minutes running on a single thread on a regular workstation.

\section{N-body simulation and H-Spine catalog}\label{sec:simulation}

The analysis presented here is based on the Illustris TNG100 simulation \citep{Nelson19,Marinacci18,Naiman18,Nelson18,Springel18}, a cosmological hydrodynamical simulation of galaxy formation enclosed in a box of 75$h^{-1}$Mpc of side. The cosmological parameters used in the simulation are $\Omega_m=0.3089, \Omega_\Lambda=0.6911, \Omega_b=0.0486, h=0.6774, \sigma_8=0.8159$. It is important to note that in the work presented here we did not make use of the publicly available TNG100 simulation datasets. Instead we constructed a hierarchical-space by running a set of dark matter-only resimulations based on the TNG100 initial conditions. In the rest of this section we describe these resimulations.

\begin{figure*}
  \centering
  \includegraphics[width=0.99\textwidth,angle=0.0]{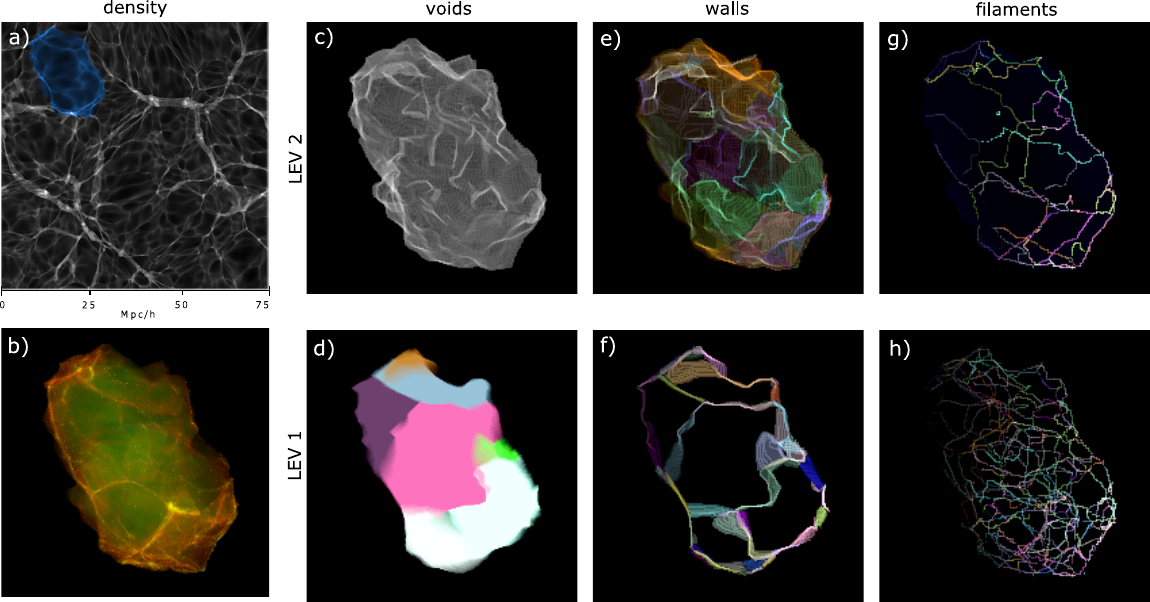}
  \caption{Hierarchy of structures inside the largest void in TNG100, Void 0. for clarity we show only two levels of the hierarchy. 
  The left column shows {\bf a)} the location of Void 0 in the simulation box, {\bf b)} a volume rendering of the interior of Void 0 . The following columns show: 
  {\bf c)} the boundaries of Void 0 and its subvoids {\bf d)}. Each void is marked with a different color. Sub-voids are shown as a volume-rendered thin slice. 
  Walls enveloping Void-0 are shown in panel {\bf e)} and sub-walls at the interior of Void-0 are shown in a thick slab cut through the center of the void in panel {\bf f)}. Note that each individual wall (sub-wall) has a different color. 
  Filaments enveloping Void 0 are shown in panel {\bf g)} and sub-filaments criss-crossing the interior of Void 0  are shown in panel {\bf h)} as colored threads.}
  \label{fig:void_decomposition}
\end{figure*}

\subsection{Hierarchical space construction}

We constructed a hierarchical space with 3 levels (LEV-0, LEV-1 and LEV-2) corresponding to a cut scale of 1, 2 and 4 \mpc respectively, at the initial conditions. Each level was computed by evolving its corresponding smoothed initial conditions file\footnote{TNG100 parameter file kindly provided by Dylan Nelson.} following the procedure described in Sec. \ref{sec:hierarchical-spaces}. The base initial conditions file was generated with the NGenIC code \citep{Springel05} using the same parameter file as the original TNG100 simulation but with a resolution of $256^3$ particles. 

Each of the smoothed initial conditions was evolved to $z=0$ using the Gadget-2 code \citep{Springel05}. From the final snapshots we computed density fields on a $512^3$ grid using the Lagrangian Sheet density estimation and interpolation method \citep{Abel12, Shandarin12} \footnote{the code can be found at https://github.com/miguel-aragon/Lagrangian-Sheet-Density-Estimator}. The density fields were further cleaned with a median filter and a $\sigma=1.5$ voxel Gaussian smoothing in order to prevent spurious oversegmentation (see Appendix \ref{sec:hspace_n_body}).

\begin{figure*}
  \centering
  \includegraphics[width=0.9\textwidth,angle=0.0]{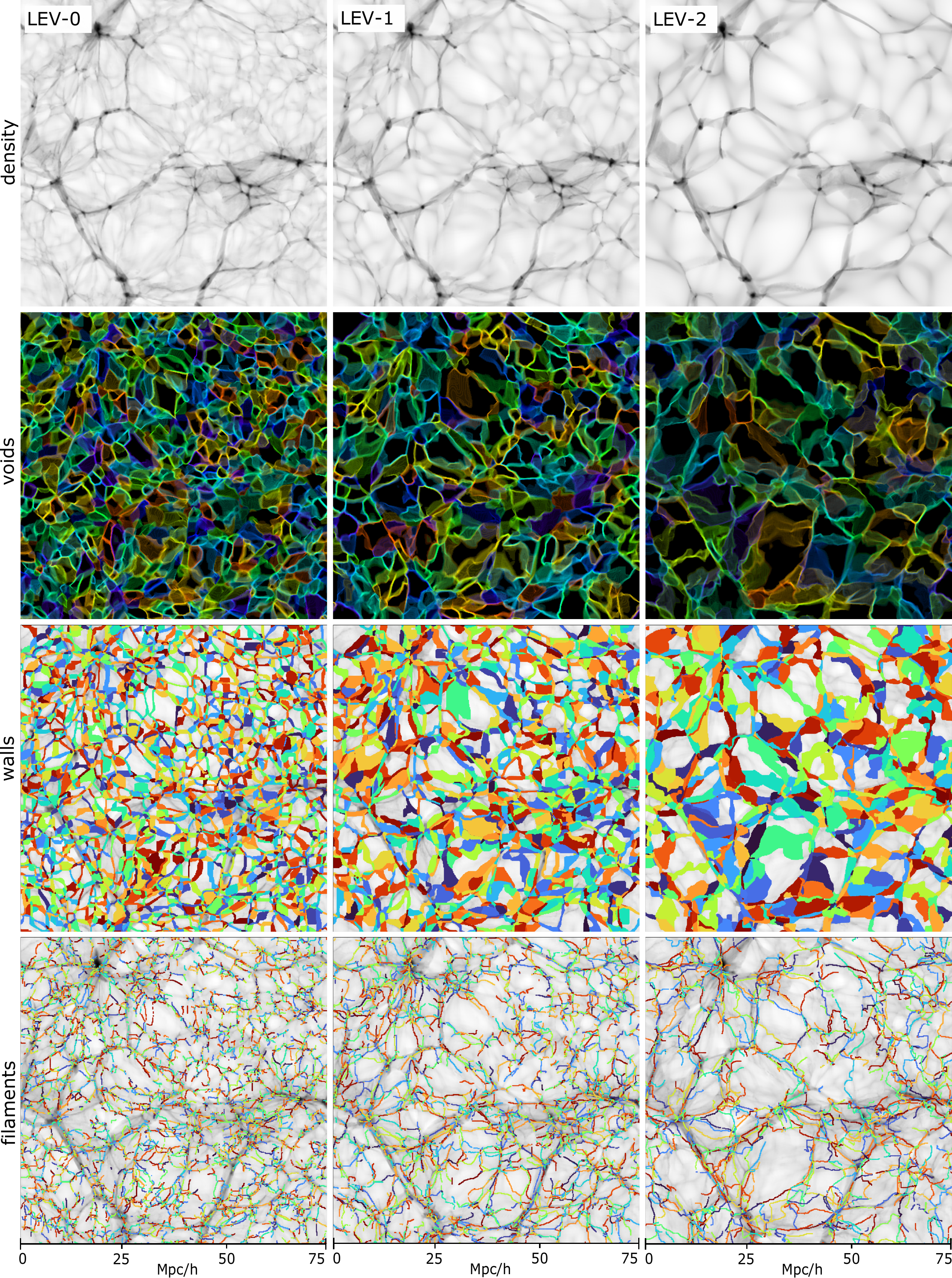}
  \caption{From top to bottom: density field, catalogue of voids, walls and filaments across a thick slice of the simulation box. From left to right we show the results for LEV-0,1,2 respectively. Individual voids, walls and filaments are colored according to the unique id in their respective catalogue.}
  \label{fig:void_catalog}
\end{figure*}

\section{The hierarchical cosmic web}

\subsection{Hierarchical deconstruction of a large void}

Following the spine formalism we identified voids, walls and filaments and their catalogs at the three hierarchical levels  described in the previous paragraph. The generated catalogs consist of sets of voxels labeled with their parent void, wall or filament unique identifier in a similar way as halo catalogs in N-body simulations consist of labeled particles (see Appendix \ref{app:examples}).\\

Figure \ref{fig:void_decomposition} shows the hierarchical decomposition of the largest void in TN100, Void 0, identified at LEV-2. The void is highlighted in a slice across the simulation box in panel a), the internal structure of the void can be seen in a volume rendering of the density field inside the boundaries of the void (panel b). The volume rendering shows a rich substructure inside the void, several sub-filaments and sub-walls cross its interior, some extending from side to side of the void. Void 0 is shown as a grey membrane in panel c) and its sub-voids are shown in panel d) where we can see several sub-voids completely filling up the volume of Void 0. Note that the boundaries of voids in LEV-N are a subset of the boundaries of voids in LEV-(N-1). Walls and filaments in LEV-0 form a watertight membrane enclosing Void 0. The walls of sub-voids in LEV-1 delineate the sub-voids in panel d). Note that since our method produces wall catalogues we are able to identify and color individual walls. Finally, filaments in LEV-2 form a net surrounding Void 0, again we stress that the Spine method provides a catalogue of individual filaments, not just voxels labeled with their morphology. The filaments form a network delineating the boundaries of walls. The network of filaments in LEV-1 is too complex to fully appreciate its connectivity in the figure. \footnote{An interactive virtual reality visualization of sub-filaments can be seen at: https://serious-debonair-witness.glitch.me}

\section{General properties of the hierarchical cosmic web}

We begin this section by providing a qualitative description of the network of voids, walls and filaments in hierarchical space. 
Figure \ref{fig:void_catalog} shows the network of voids identified at three hierarchical levels and their corresponding density field. The network of voids at each level delineates the visually salient voids in the density field with increasingly larger void sizes for increasing hierarchical level. The use of a hierarchical space conserves the anisotropies in the density field, large voids seem to have more complex shapes while smaller voids tend to be more spherical. 
Some voids appear unchanged across adjacent levels in the hierarchy. This is expected since voids do not gradually increase in size across hierarchical space, instead voids are quasi-stable structures in hierarchical space that can remain stable and undergo drastic change after some critical smoothing scale. This will be studied in a future work. 

Figure \ref{fig:void_catalog} shows the catalog of walls across a thick slab of the simulation. Individual walls form a continuous surface marking the boundaries of voids. In contrast pixel-based analysis like those based on the Hessian matrix produce discontinuous structures and they can not identity individual objects. Just as in the case of voids, walls increase their area with increasing level in the hierarchy as they are the membranes enclosing increasingly larger voids. Also walls increase their complexity following the shape of their enclosed voids as we move higher in the hierarchy. 

The catalog of filaments for three hierarchical levels is shown in Figure \ref{fig:void_catalog}. As in the case of voids and walls, filaments increase their length as we move higher in the hierarchical space. Filaments in LEV-0 are short and mostly straight and correspond to the bridges between nearby haloes, groups and clusters, this has been observed in manually selected filaments \citep{Colberg05}. The complexity of filaments increases with hierarchical level. Filaments in LEV-2 are long and curved. Spine filaments in LEV-2 trace dense filaments in the density field. Filaments identified at lower levels in the hierarchy are sub-filaments with lower density.

\begin{figure}
  \centering
  \includegraphics[width=0.45\textwidth,angle=0.0]{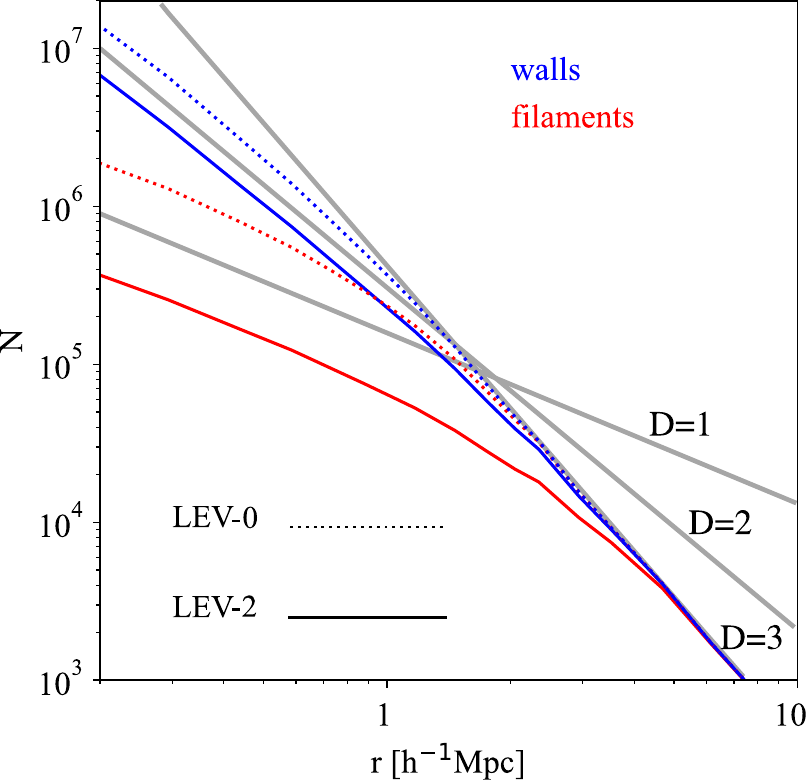}
  \caption{Minkowski–Bouligand dimension computed for filaments and walls in LEV-0,1,2 (dotted, dashed and solid lines respectively). For reference we show dimensions D=1,2 and 3 as thick gray lines. For clarity we only show the Minkowski–Bouligand dimension for LEV-0 and LEV-2. The curve corresponding to LEV-1 is an intermediate case.}
  \label{fig:fractal}
\end{figure} 

\subsection{Dimensionality of walls and filaments}

A quantitative measure of the characteristic scale of the network of walls and filaments can be given by the Minkowski–Bouligand dimension or box counting dimension which measures the fractal dimension of an object across scales. In our case we can apply it to study the dimensionality of the network of walls and filaments (the network of voids is 3-D at all scales) and identify any changes that may indicate a transition in scale which would be shown as a change in its fractal dimension. 
The Minkowski–Bouligand dimension is defined as:

\begin{figure*}
  \centering
  \includegraphics[width=0.99\textwidth,angle=0.0]{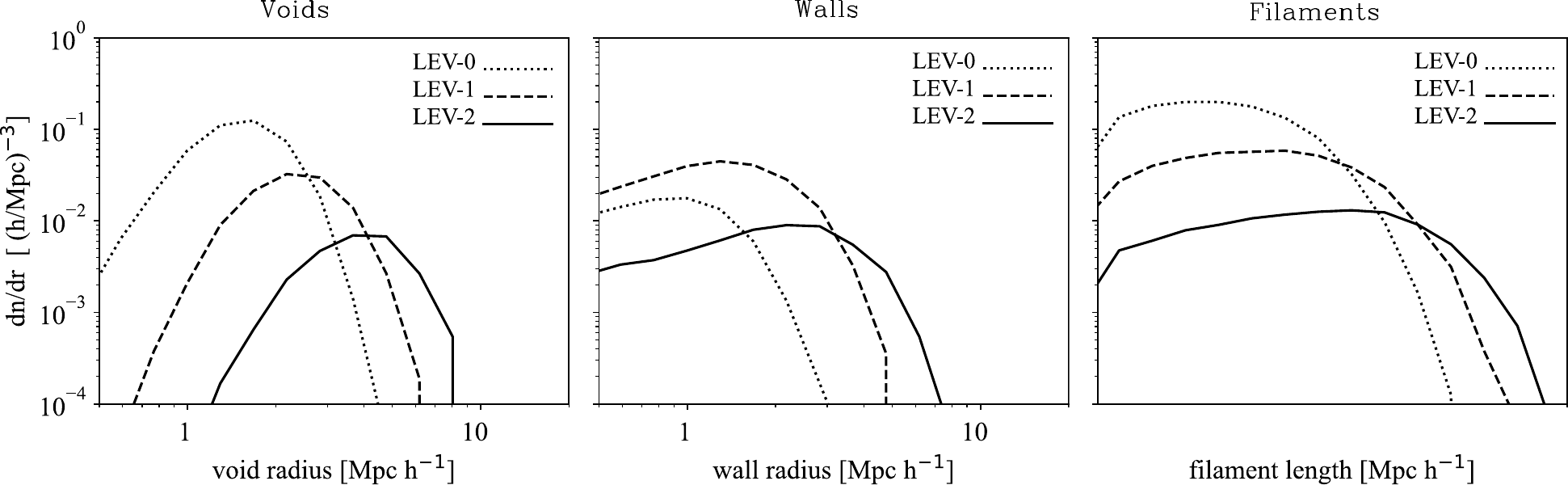}
  \caption{Distribution of void equivalent radius, wall equivalent radius and filament length in three hierarchical levels used in this work.}
  \label{fig:size_distribution}
\end{figure*}

\begin{equation}
\textrm{Dim}_{box}(S) = \lim_{\epsilon \to 0} \frac{\textrm{Log}N(\epsilon)}{\textrm{Log}(1/\epsilon)}.
\end{equation}

\noindent Where $N(\epsilon)$ is the number of boxes of infinitesimal size $\epsilon$ required to fully cover the set of voxels belonging to the filament or wall network. In practice, we divide the simulation box into subboxes of size $S$ and count the number $N(S)$ of subboxes that contain at least one voxel labeled as filament or wall. By repeating this evaluation for several box sizes, we obtain an estimate of the Minkowski-Bouligand dimension. The Minkowski–Bouligand dimension, shown as a plot of $N(S)$ versus box size $S$ in logarithmic scale, corresponds to the slope of the resulting curve. 

Figure \ref{fig:fractal} shows the Minkowski–Bouligand dimension for walls and filaments in hierarchical levels LEV-0, LEV-1 and LEV-2. The network of both filaments and walls behaves as a three dimensional structure at large scales and changes to 1D and 2D respectively at small scales as the network of walls and filaments transition from being an interconnected network to being a set of individual structures. This is more clear for filaments in LEV-2 where we see a departure from 1D to 3D around 5 \mpc. The point of departure correlates with the linear-regime smoothing scale in each hierarchical level. In the case of walls the transition point is not so sharp as in the case of filaments but still there is a clear correlation between transition scale and linear-regime smoothing length. 

\subsection{Size distributions}

A more detailed view of the size distribution of voids, walls and filament is presented in Fig. \ref{fig:size_distribution} where we show the distribution of sizes of voids, walls and filaments in the three hierarchical levels studied in this work.

\subsubsection{Void equivalent radius}

The distribution of sizes of voids is shown in Fig. \ref{fig:size_distribution} (left panel, see also \citet{Colberg05,Platen08} for single-scale analysis). We compute the equivalent radius of a void, i.e. the radius of a spherical void of the same volume as a given void as:

\begin{equation}
 r_{\textrm{\tiny void}} = \left (\frac{3}{4 \pi} V_{\textrm{\tiny void}} \right )^{1/3}.
\end{equation}

\noindent Where $V_{\textrm{\tiny void}}$ is the sum of the volume of the voxels inside the void. Given the discrete nature of the voxels composing the voids there is a small discretization effect only noticeable for very small objects. The equivalent radius of voids increases with level in the hierarchy with peaks in the distribution at $\sim 1.5,2.5 $ and $4$ \mpc for LEV-0, 1, and 2 respectively. The linear-smoothing applied to create the hierarchy acts as low-pass filter removing primordial fluctuations that would result in voids smaller than the smoothing length being suppressed. One would expect to see a cutout at void radius approximately smaller than the smoothing length corresponding to the level in the hierarchy and voids of large sizes limited only by the simulation box size. However, we observe voids in a relatively narrow radius range for each level in the hierarchy. The oversegmentation of the watershed transform (i.e. the segmentation of the image in regions smaller than the features one seeks to identify) reflects small-scales present in the power spectrum. Oversegmentation  effectively creates a band-pass filter in void-size as shown in the radius distributions. See for instance Fig. \ref{fig:void_merge} in which the larger voids in the simulation are decomposed into smaller voids. In the Hierachical Spine method the smaller voids correspond to real structures and not spurious detections as it occurs with the usual oversegmentation problem.

\subsubsection{Wall radius}

\rev{Figure \ref{fig:size_distribution} (center panel) shows the distribution of wall equivalent radius for the three hierarchical levels.} The equivalent radius of walls is defined as the radius of a circle with the same area as the wall. We computed the area of walls by first aligning the plane of each wall with the xy plane and then computing the convex hull encompassing the voxels forming the wall (see appendix \ref{app:wall-area} for details). The distribution of wall sizes shows a characteristic scale at the location of the peak in the distribution at $\sim$ 1, 2, and 3 \mpc for LEV-0,1,2 respectively. As a practical application of this result we can consider the distribution of wall sizes across the hierarchy and compare to the scale of the Local Sheet \citep{Tully08,Aragon11,McCall14}, with a radius of $\sim$ 4 Mpc is sitting near the peak in the wall size distribution at LEV-2, suggesting that our cosmic environment may have originated from a $\sim$ 4 Mpc fluctuation (see also \citep{Einasto11}.

\subsubsection{Filament length}

Figure \ref{fig:size_distribution} (right panel) shows the distribution of filament lengths for the three hierarchical levels. The filament lengths was computed by fitting a degree-3 polynomial to the voxels forming a given filament (see app. \ref{app:filament-length} for details). The polynomial fit is sufficient to closely follow most of the filaments except a small fraction of filaments with complex shapes for which their length would be under estimated. Interestingly the peak in the distribution of filaments is located at higher values than the peak in the distribution of both voids and walls. This is simply because we are comparing  filament length with equivalent radius of voids and walls and not with their diameter. Filaments can border a wall or a void (see Fig.  \ref{fig:void_decomposition}) and easily cover the length of one diameter as suggested by the high-length tail in the filament distribution located roughly at twice the corresponding radius of voids and walls.

\subsection{Density distributions}


\begin{table}
\begin{tabular}{ l c c}
\hline
Structure & mean density (LEV-2) & median density (LEV-2) \\ \hline
Void      & 0.51  &  0.17 \\
Wall      & 1.52  &  0.45 \\
Filament  & 5.59  &  1.20 \\
\end{tabular} 
\caption{Mean and median values of the density field inside voids, walls and filaments identified at LEV-2.}\label{tab:densities}
\end{table}

\begin{figure*}
  \centering
  \includegraphics[width=0.98\textwidth,angle=0.0]{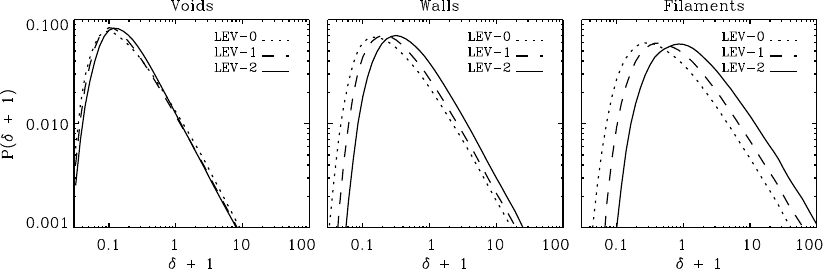}
  \caption{Density distribution inside voids, walls and filaments defined in the three hierarchical levels used in this work.}
  \label{fig:density_distribution}
\end{figure*}

Figure \ref{fig:density_distribution} shows the distribution of densities inside voxels labeled as voids, walls and filaments in the three hierarchical levels used in this work. The Spine method produces structures with a thickness of $\sim$one voxel which maps the cores of both filaments and walls, producing an artificially high density distribution. In order to measure the density distributions in a more physically meaningful way we apply the dilation morphological operator to the wall and filament masks in a recursive way 3 times. This produces walls and filaments with thickness of $\sim$1 \mpc (see \citet{Colberg05}).

The density distribution in voids, walls and filaments has a roughly log-normal character, consistent with the non-linear evolved cosmic web (see \citet{Neyrinck09}). The density distribution inside voids is very similar between the three hierarchical levels, with a peak at $\delta+1 = 0.1$. LEV-2 voids are slightly denser than voids at LEV-0,1 as the result of the contribution of the nested tenuous cosmic web inside voids (sub-boids, sub-walls, sub-filaments, etc.). Voids at lower levels in the hierarchy contain less substructure and their substructure is less dense. 

\rev{The density distribution increases with both hierarchical level and LSS morphology following the order: voids, walls and filaments, according to the stages of gravitational collapse described in Sec \ref{sec:grav_collapse}. Spine elements increase their density at higher levels in the hierarchy. Filaments defined by the largest scales at LEV-2 are also the most prominent (i.e. the densest) in the LSS, on the other hand, filaments identified at LEV-0 can be found inside voids and crossing walls and have much lower densities.} Note that the space between the peaks in the density distributions is also more pronounced in the case of the filaments compared to walls and voids in that order respectively.

We note that the mean density inside voids, walls and filaments at LEV-2, which corresponds to the largest structures in the hierarchy, is remarkably close to the commonly assigned values for these structures.

\section{Conclusions}

We presented the Hierarchical Spine (H-Spine) method for the identification of voids, walls and filaments in a hierarchical fashion. We show that the H-Spine method can successfully identify voids, walls and filaments as well as their sub-structures within the hierarchy. The cosmic web, being a hierarchical system, can only be properly described with a hierarchical prescription as the one presented here. Furthermore H-Spine is model agnostic, making no assumptions on the underlying cosmology. However, models with modified gravity such as $f(R)$ may be affected by the smoothing in the initial conditions at small scales. In such cases one can choose a different method to construct a hierarchical space such as Gaussian smoothing or thresholding as discussed in Sec. \ref{sec:hierarchical-spaces}.

We introduced a physically motivated prescription to identify walls and filaments in terms of voids, inspired by the order of gravitational collapse and intrinsic connectivity of the elements in the cosmic web (see Sec. \ref{sec:spine}). The identification of structures is performed in a hierarchical fashion and the formalism for this is also introduced. The hierarchical identification is based on hierarchical spaces which are a generalization of scale spaces in which the scale operator is applied in the linear regime, before Fourier mode mixing by gravitational evolution.

The H-Spine method is unique in its ability to produce catalogs of voids, walls and filaments and their hierarchical relations. We presented size distributions of voids walls and filaments. The distribution of individual wall sizes is presented here for the first time as far as the author knows. We also presented the distribution of densities in voids, walls and filaments and show that the mean values of the density at the top of the hierarchy (LEV-2) are close to commonly assigned values.

The identification of cosmic structures in hierarchical spaces is not restricted to topological analysis tools like the H-Spine presented here. Hierarchical spaces can be also analyzed with local gradient LSS methods as done with the MMF-2 method introduced in \citet{Aragon13b}. However, local gradient and other local LSS analysis tools can only benefit from the cleaner density fields provided by the initial condition smoothing used to generate the hierarchical space and can only partially expose hierarchical relations between the elements of the cosmic web.

The study of the cosmic web as a hierarchical system can give us new insights on galaxy formation processes as shown in the dependence of spin alignment on hierachical level \citep{Aragon14} and the properties of galaxies in thin filaments inside voids \citep{Alpaslan14}. The formalism presented here allows us to explicitly describe the connectivity of the elements of the cosmic web across the hierarchy and create a cosmic web hyper-graph, providing a new and powerful language to describe the cosmic web and understand its relation to galaxy formation processes (Aragon-Calvo in preparation).

\section{Data availability}
The H-Spine code will be made publicly available after the paper has been accepted for publication. The data used in this paper can be provided after a reasonable request is made to the author.

\section{Acknowledgements}
The author acknowledges support for the DeepVoid project and this work from grant 62177 from the John Templeton Foundation. Also would like to thank Dylan Nelson for kindly providing the TNG100 parameter file used for generating the initial conditions. 

\bibliography{refs} 
\bibliographystyle{mn2e}   

\appendix

\section{Creating hierarchical spaces}\label{sec:hspace_n_body}

\begin{figure}
	\centering
	\includegraphics[width=0.49\textwidth]{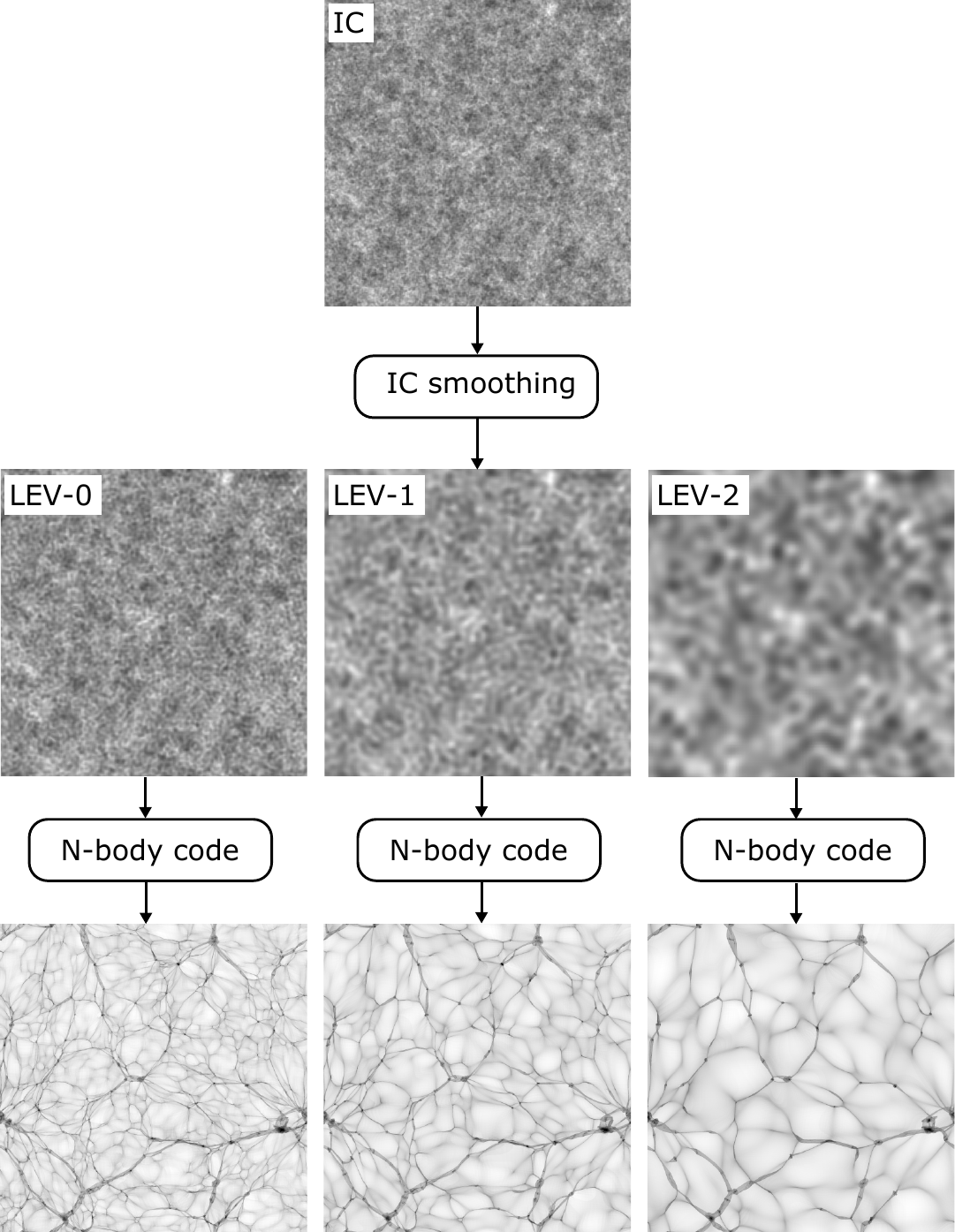}  
	\caption{ \rev{Construction of a hierarchical space with 3 levels, starting with the full initial conditions (top) three levels are generated by sharp-$k$ smoothing (middle) and independently evolved to $z=0$ (bottom panels).} }
	\label{fig:constructing_hierarchical_space}
\end{figure}

In practice we generate a hierarchical space as follows:

\begin{itemize} 
\item[1)] Generate initial conditions file.
\item[2)] Apply a series of low-pass sharp-$k$ filters to the initial conditions at the scales of interest.
\item[3)] Evolve each initial conditions file independently using an N-body code.
\end{itemize}

\noindent \rev{A hierarchical space thus requires running several versions of the target simulation as illustrated in Fig. \ref{fig:constructing_hierarchical_space}. In practice this can be achieved at significantly lower resolution (and computational cost) compared to the target simulation. In this work the target simulation is Illustris-TNG100 which contains $1820^3$ particles in a $75$ \mpc box at the highest resolution. The simulations used to generate the hierarchical space have $512^3$ particles each, corresponding to a mean interparticle separation of $0.14$\mpc which is sufficient to sample the filaments joining galaxies and groups of galaxies. Each of the $512^3$ particle simulations took less than two days to run on a workstation with 32 cores. Furthermore, the use of approximate codes such as L-PICOLA \citep{Howlett15} can speedup the process by several orders of magnitude.
}

\section{Illustrative examples}\label{app:examples}

\begin{figure}
  \centering
  \includegraphics[width=0.3\textwidth,angle=0.0]{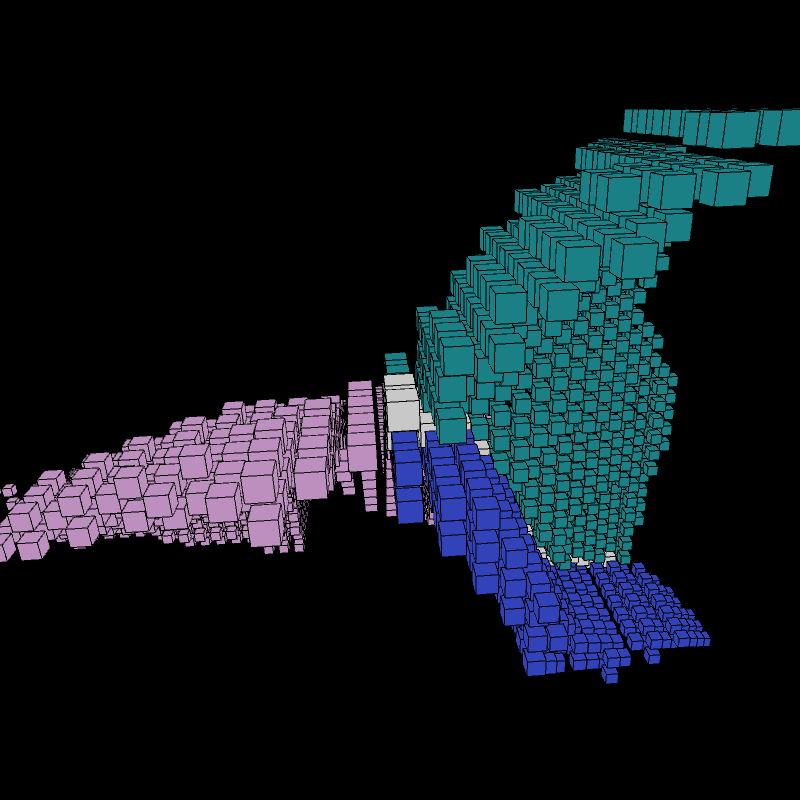}
  \includegraphics[width=0.3\textwidth,angle=0.0]{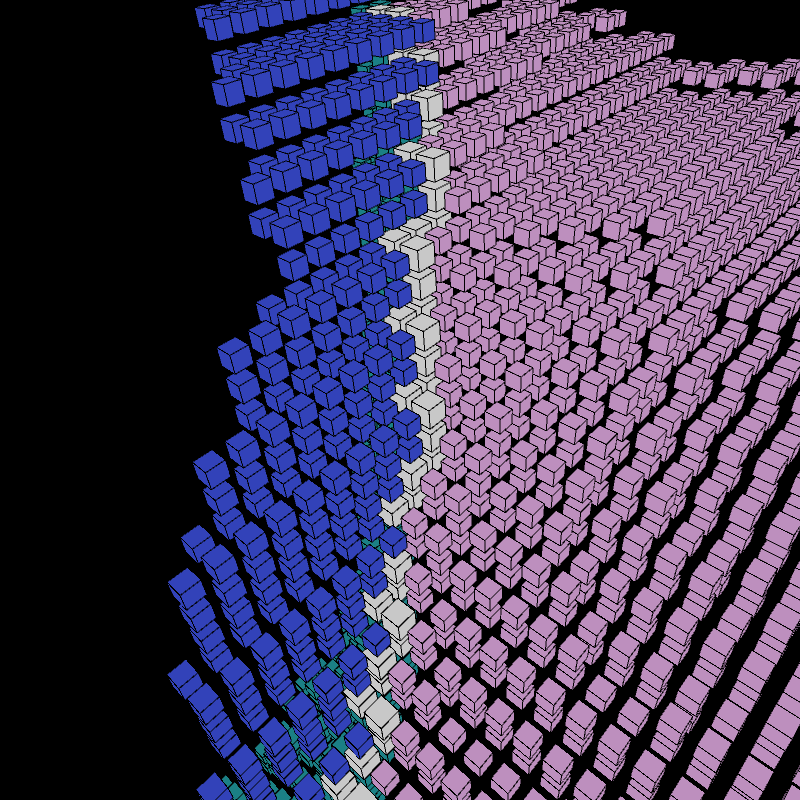}
  \includegraphics[width=0.3\textwidth,angle=0.0]{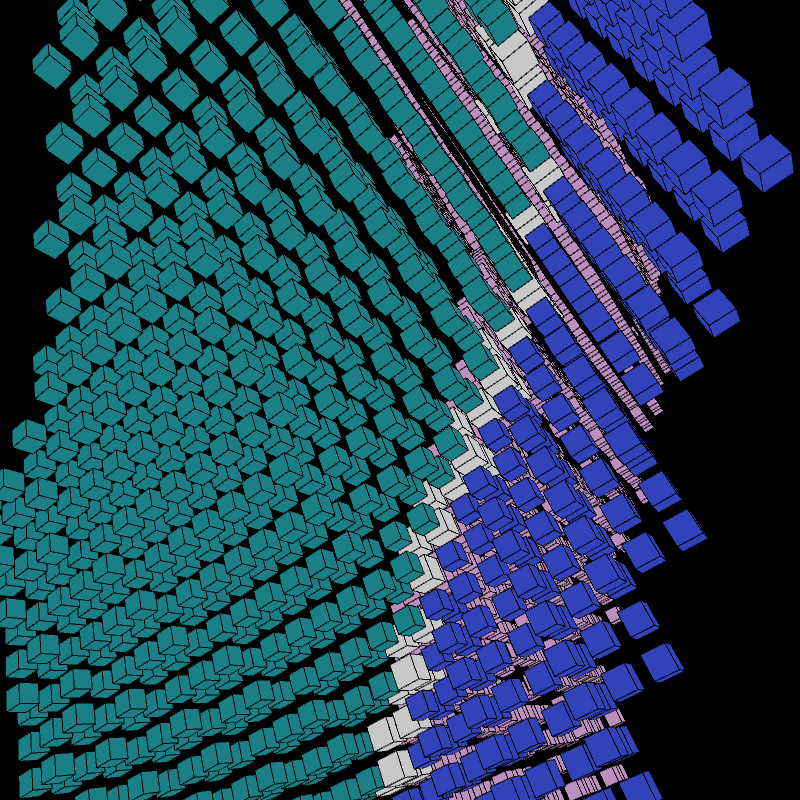}
  \caption{Three views (front view and two side views) of a filament and its 3 adjacent walls . White cubes correspond to the voxels shaping the filament and colored cubes correspond to the voxels of its adjacent walls. For clarity cubes/voxels are displayed at a smaller size than their size in the grid.  Compare this to Fig. \ref{fig:cosmic-web-connectivity} (central panel) where we show a toy model of a similar filament-wall configuration.}
  \label{fig:filament-proj}
\end{figure}

\begin{figure}
  \centering
  \includegraphics[width=0.3\textwidth,angle=0.0]{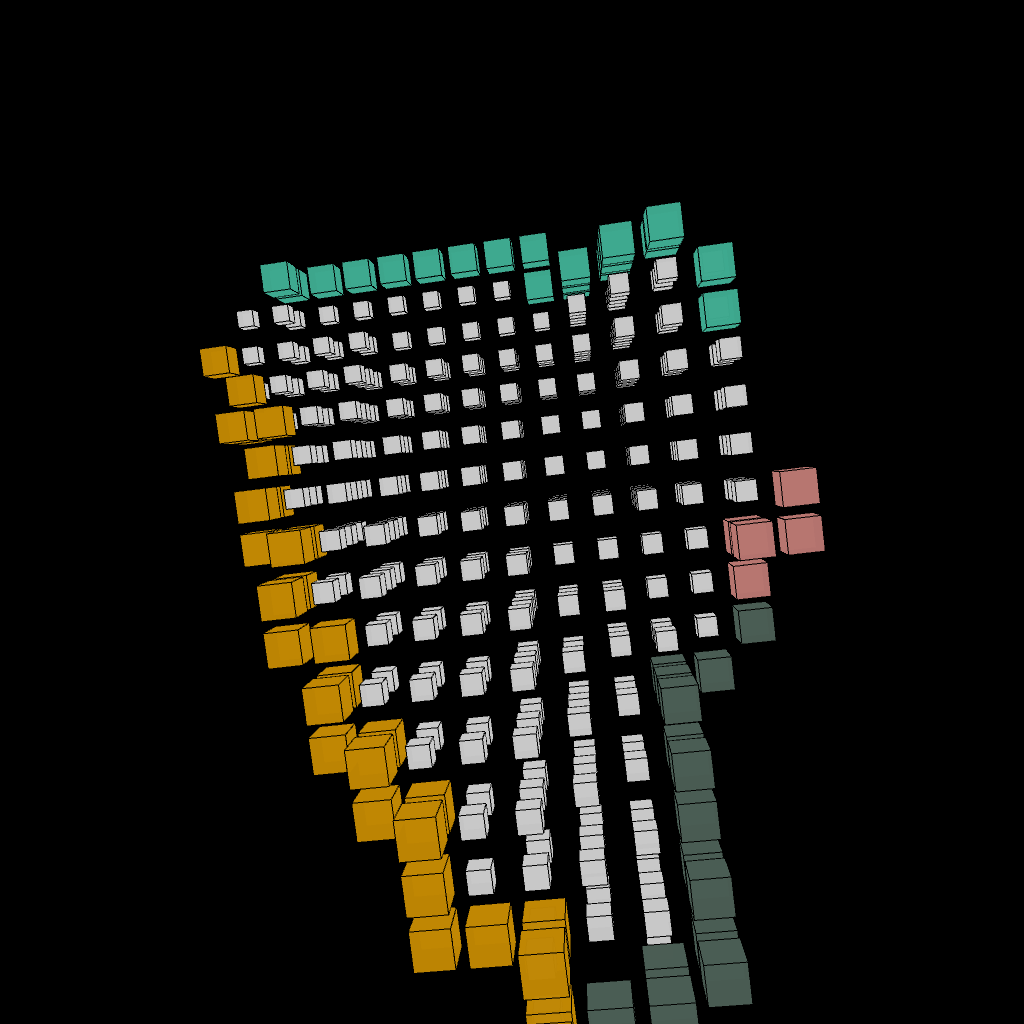}
  \caption{Face-on view of a wall and its adjacent filaments. White voxels correspond to the wall and colored voxels to its adjacent filaments. For clarity voxels are displayed at a smaller size than their size in the grid.}
  \label{fig:wall-proj}
\end{figure}

\begin{figure}
  \centering
  \includegraphics[width=0.3\textwidth,angle=0.0]{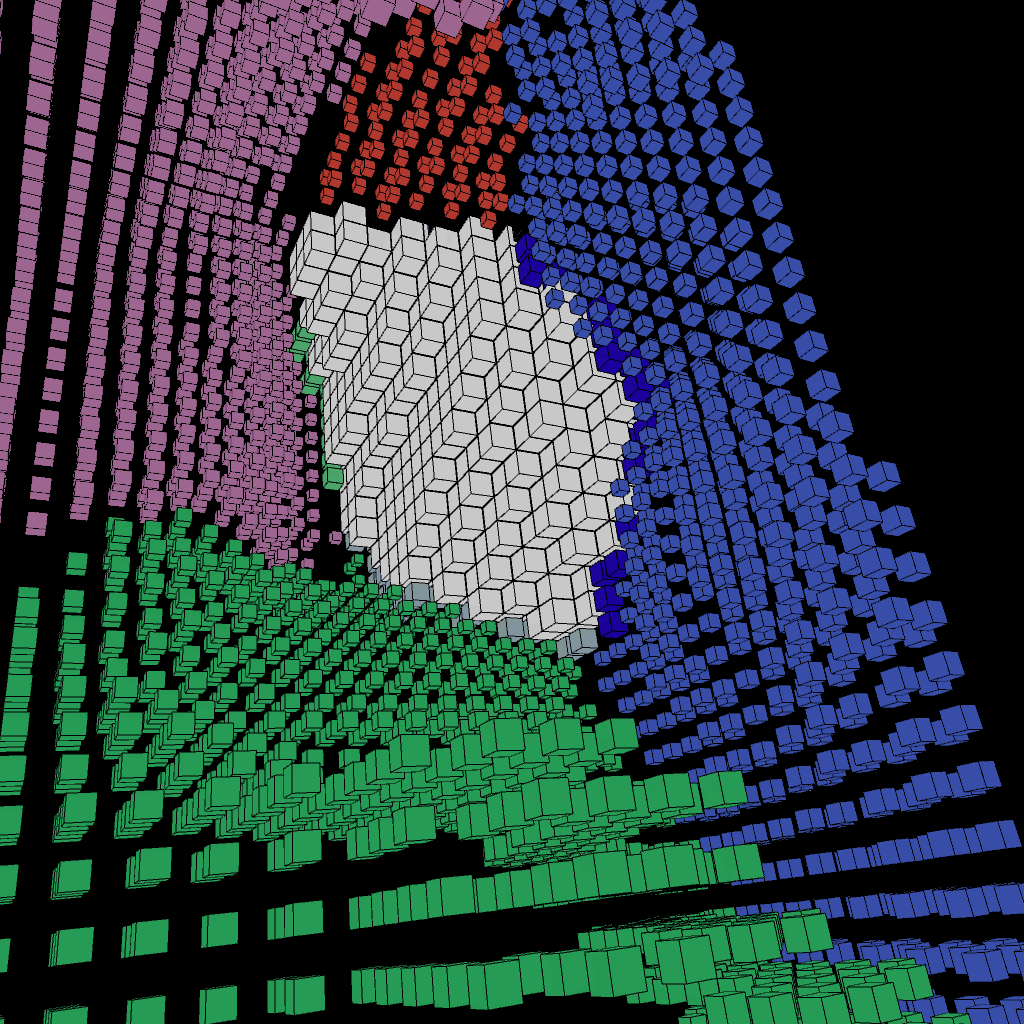}
  \caption{A central wall and its adjacent walls. White voxels correspond to the central wall and colored voxels to its adjacent walls. We also show the adjacent filaments to the central wall but given the complexity of the figure they are not fully visible. For clarity voxels are displayed at a smaller size than their size in the grid.}
  \label{fig:wall-adj}
\end{figure}

In this section we present some cases of filaments and walls that highlight the connectivity of the elements of the cosmic web extracted with the new Spine formalism.
Figure \ref{fig:filament-proj} shows three different projections of a filament and its adjacent walls. There are three walls adjacent to this filament that run through most of its length. Every pair of adjacent walls are part of the envelope surrounding a void that is both adjacent to the filament and its adjacent walls. Figure \ref{fig:wall-proj} shows a wall and its adjacent filaments. There are 4 filaments connected to this particular wall and they surround and define its boundaries.
Finally, Fig. \ref{fig:wall-adj} shows a wall and its adjacent neighbouring walls. Filaments, being the intersection of walls, define their connectivity. Several filaments can be seen at the intersecting edges of adjacent walls. Together these walls form part of the envelope of a void whose center is located towards the viewer.

%
\section{Computing wall areas}\label{app:wall-area}

\begin{figure}
  \centering
  \includegraphics[width=0.5\textwidth,angle=0.0]{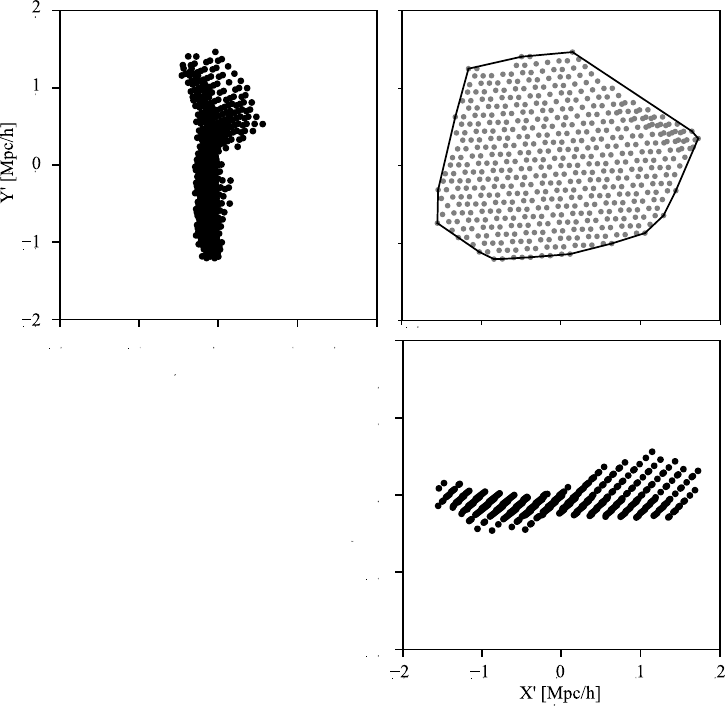}
  \caption{Wall area estimation. We show three orthogonal projections of the voxels forming a wall after they have been rotated and aligned with the $xy$ plane. The convex hull, used to compute the area of the wall, is shown as a polygon enclosing the point set in the top-right panel. Voxels, represented as points, are shown as dots.}
  \label{fig:wall-model}
\end{figure} 

Computing the area of a wall is a challenging problem as walls are not perfectly flat structures with smooth borders. We now describe an approximate method that can be used to estimate the area of walls extracted with the H-Spine method.

We start by converting the position of the voxels corresponding to a given wall to points in space according to their location in the data cube. The voxels in the wall are a relatively flat point cloud, we use Principal Component Analysis to find the main axis of the cloud. The wall is rotated in order to align its plane with the $xy$ plane as shown in Fig. \ref{fig:wall-model}. We then find the convex hull enclosing the projected points and compute the area of the convex hull. As Fig. \ref{fig:wall-model} shows this approximation works well although it slightly underestimates the area of warped walls. Based on visual inspection of a large number of walls we find that most walls are relatively flat and have smooth convex borders. The algorithm presented here is easy to implement and fast, allowing us to compute areas for all walls in the simulation in a couple of minutes on a regular workstation.

%
\section{Computing filament lengths}\label{app:filament-length}

\begin{figure}
  \centering
  \includegraphics[width=0.23\textwidth,angle=0.0]{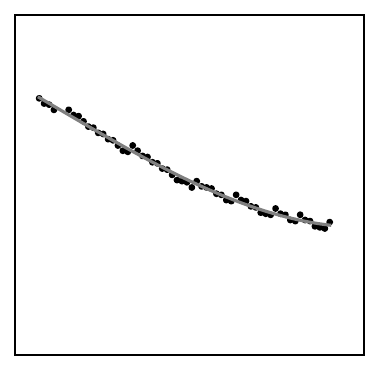}
  \includegraphics[width=0.23\textwidth,angle=0.0]{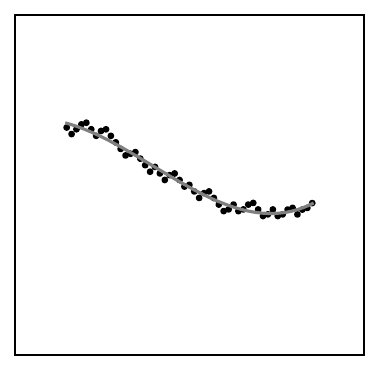}
  \includegraphics[width=0.23\textwidth,angle=0.0]{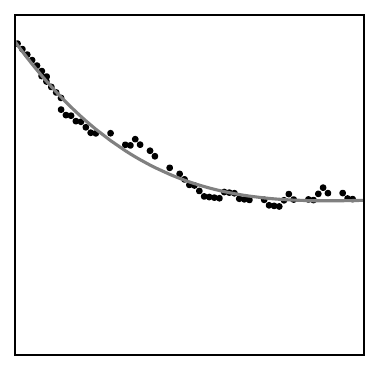}
  \includegraphics[width=0.23\textwidth,angle=0.0]{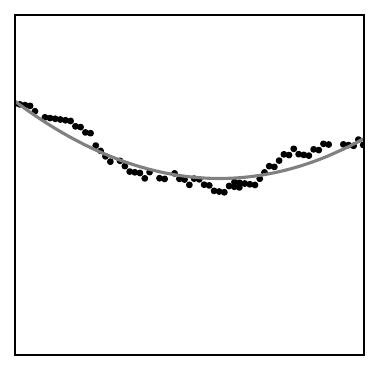}
  \caption{Filament length estimation. We show four different filaments and their fitted degree-3 polynomial (solid gray line). Voxels, converted to points, are shown as black dots.}
  \label{fig:filament-model}
\end{figure} 

In order to estimate the length of filaments extracted with the H-Spine method we used a polynomial fit of degree 3 to the voxels inside a given filament. As in the case of walls the voxels in filaments were first converted to points. Every filament was then translated to the origin and fitted to the polynomial. After visual inspection of a large number of filaments we found that a polynomial of degree 3 is sufficient to follow the large majority of filaments. Only very few large filaments with complex shapes could not be well fitted. In such cases the filament length was underestimated. However given the low number of such cases this did not affect our analysis. A more sophisticated algorithm could be implemented with a polynomial of different degree for each individual filament. Such analysis is out of the scope of the present work but will be addressed in future studies.

\end{document}